\documentclass[12pt]{iopart}
\usepackage{graphicx}
\usepackage[export]{adjustbox}
\usepackage{subfig}
\usepackage{setstack}
\usepackage{cite}
\usepackage{amsfonts}
\usepackage{tcolorbox}
\usepackage{braket}
\usepackage{subfig}
\usepackage{url}

\begin{document}

\title{Supervised learning with a quantum classifier using a multi-level system}

\author{Soumik Adhikary$^1$, Siddharth Dangwal$^2$ \& Debanjan Bhowmik$^3$}

\address{$^1$ Department of Physics, Indian Institute of Technology Delhi, New Delhi - 110016, India. \\
$^2$ $^3$ Department of Electrical Engineering, Indian Institute of Technology Delhi, New Delhi - 110016, India}

\ead{\mailto{$^1$soumikadhikary@physics.iitd.ac.in},\mailto{$^2$Siddharth.Dangwal.ee117@ee.iitd.ac.in} and \mailto{$^3$debanjan@ee.iitd.ac.in}}

\vspace{10pt}
\begin{indented}
\item[] August 2019
\end{indented}

\begin{abstract}
We propose a quantum classifier, which can classify data under the supervised learning scheme using a quantum feature space.  The input feature vectors are encoded in a single qu$N$it (a $N$ level quantum system), as opposed to more commonly used entangled multi-qubit systems. For training we use the much used quantum variational algorithm- a hybrid quantum-classical algorithm, in which the forward part of the computation is performed on a quantum hardware whereas the feedback part is carried out on a classical computer. We introduce ``single shot training" in our scheme, with all input samples belonging to the same class being used to train the classifier simultaneously. This significantly speeds up the training procedure and provides an advantage over classical machine learning classifiers. We demonstrate successful classification of popular benchmark datasets with our quantum classifier  and compare its performance with respect to some classical machine learning classifiers.  We also show that the number of training parameters in our classifier is significantly less than the classical classifiers.
\end{abstract}

%
\vspace{2pc}

\section{Introduction}
Quantum computation, in recent times, has gained immense popularity owing to the large number of interesting applications associated with it \cite{Bennett93, Bennett84, Ren17, nielsenbook}. A common feature, which all of these applications share, is the use of various non-classical attributes of a quantum state as their resources.  While some of these tasks do not even have a classical counterpart \cite{Bennett93}, the others show a clear  quantum advantage over their classical counterparts \cite{Grover96, Ekert91}. 

A new application that has recently emerged in the  area of quantum computing is machine learning \cite{Biamonte17}. Machine learning has long been studied in the classical setting and has proved to be highly effective for data classification and regression problems \cite{LeCun,bishop06}. It is therefore pertinent to inquire if there is a scope of improvement if the principles of quantum mechanics are also used in these machine learning problems. A number of  quantum algorithms has already been proposed to this end \cite{havliv19,Schuld19,wan17,Farhi18,Lloyd18}.

A class of quantum machine learning protocols, which  has gained much prominence of late, employ the so called variational algorithms \cite{Farhi18,McClean16,Mitarai18}. These are hybrid quantum-classical algorithms and have an added advantage of being implementable on near term quantum computers. At this stage, it is prudent to mention that there exists a large number of learning techniques- supervised, unsupervised, reinforcement, etc. \cite{russell16, hinton99, sutton98}. For the purpose of the paper we shall restrict our discussion to supervised learning only.

Consider a dataset $\mathcal{S} = \{ ({\bf x}, f({\bf x}))\}$. Each entry in $\mathcal{S}$ is an ordered pair consisting of an input vector ${\bf x}$ of arbitrary dimension  and its associated label $f({\bf x})$, corresponding to the class the input belongs to. Thus, $f$ maps each input vector to a label, from a set of labels: $\mathcal{L} = \{ l_1, l_2, \cdots, l_N\}$; $f: {\bf x} \rightarrow \mathcal{L}$. Each label corresponds to an output class. There are $N$ classes in total. The objective of supervised learning is to train a machine using a subset $\mathcal{T}$ chosen from the given dataset such that the machine can infer correct labels for the train set $\mathcal{T}$ as well as the test set $\mathcal{S}-\mathcal{T}$. More rigorously speaking, we require the machine to return a function $f^*$ so that $f^*({\bf x}) = f({\bf x})$ for maximum number of input vectors \cite{LeCun,bishop06}. 

A typical quantum protocol for supervised learning, that employs variational algorithms, broadly consists of three stages \cite{Schuld18}. The first stage is state preparation. Every input vector gets encoded into a quantum state. The encoding scheme by itself may vary widely \cite{havliv19,Farhi18,Schuld18}. In the second stage the quantum state gets acted upon by a set of parameterized unitary operations. All of these parameters are tunable. Finally the predicted label for a given input vector is obtained, following  a projective measurement on the transformed state. As mentioned earlier, in general, the predicted label is not always the same as the actual label, i.e, $f^*({\bf x})$ may not be equal to  $f({\bf x})$. To ensure that the two agree for most cases, the error function $E(\vert f^*({\bf x}) - f({\bf x}) \vert)$ is minimized during the training phase. This is carried out by updating the tunable parameters in the unitary operations, for every training input. Although the forward computation takes place on a quantum computer, the evaluation of the weight update needed at that iteration needs to be calculated on a classical computer using the stochastic gradient descent algorithm \cite{LeCun}. This makes the algorithm a hybrid between quantum and classical regimes \cite{McClean16}.

In this paper we  propose a new implementation of a quantum classifier, making use of the variational algorithm. As declared we shall use the supervised learning technique. Our implementation differs  from the already existing ones in a number of key ways. The most prominent among them are as follows:

\begin{enumerate}
    \item We encode the input vectors in states of a $N$-level system, as opposed to more conventionally used entangled multi-qubit states. Quantum hardware using entangled multi-qubit states are inherently noisy \cite{Preskill18}. In contrast, multi-level quantum states (with dimension as high as 27) have been prepared with fairly high precision \cite{mehul14}.
 \item We introduce a technique, which we call ``single shot training", in our scheme. This enables us to train the circuit, for all samples belonging to the same output class, at once, thus providing a speedup of the training process.
\end{enumerate}

We discuss our implementation in details in Sec. \ref{meth}. We present our results of training the classifier on popular benchmark datasets- Fisher's Iris dataset \cite{FisherIris}, Sonar dataset \cite{Sonar} and Wisconsin's Breast Cancer  (WBC) dataset \cite{WBC}  in Sec. \ref{results}. Subsequently we compare them with results from classical feed-forward Fully Connected Neural Network (FCNN) and discuss the advantages of our quantum classifier over the latter. In Section \ref{disc}, we discuss the various subtleties associated with our implementation and conclude the paper.

\section{Method}
\label{meth}

\begin{figure}
    \centering
    \includegraphics[width=0.9\textwidth]{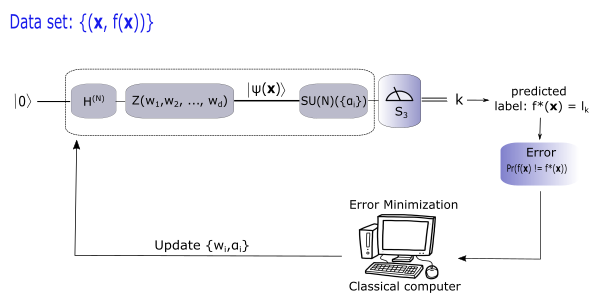}
    \caption{A schematic representation of our quantum classifier. The gates $H^{(N)}$, $Z(w_1, w_2, \cdots, w_d)$ and $SU(N)(\{\alpha_i\})$ are defined in Eq.~(\ref{encoding 1}), (\ref{genhadamard}) and (\ref{group}). $S_3$ is the diagonal matrix $Diag(1,2, \cdots, N)$ and $f^*$ is the decision function that maps ${\bf x} \rightarrow \mathcal{L}$.}
    \label{schematic}
\end{figure}

Consider once again the dataset $\mathcal{S} = \{ {\bf x}, f({\bf x}) \}$; ${\bf x} \in \mathbb{R}^d$, $f: {\bf x} \rightarrow \mathcal{L} = \{ l_1, l_2, \cdots, l_N\}$. $\mathcal{L}$ is the set of labels. As already mentioned, a quantum classifier employing variational techniques, consists of three stages. We shall describe next, how we have implemented these stages in our scheme (see Fig. \ref{schematic} for a schematic).

\subsection{State preparation}
\label{sp}

Let us start with a single qu$N$it (a system with $N$ levels)  belonging to the Hilbert space $\mathcal{H}^N$. Note that we have ensured the dimension of the Hilbert space to be equal to the cardinality of $\mathcal{L}$ i.e the total number of output classes. This is imperative to our scheme. The input vectors ${\bf x}$, will be represented in $\mathcal{H}^N$ by $\ket{\psi(\bf x)}$. The encoding scheme that we employ to obtain $\ket{\psi(\bf x)}$ is given in Eq.~(\ref{encoding 1}).

   \begin{eqnarray}
   \label{encoding 1}
 \ket{\psi(\bf x)} &= e^{i \overline{S}_3 (\sum_{j=1}^d w_j x_j)} H^{(N)} \ket{0} \\ \nonumber
   &= Z (w_1, w_2, \cdots, w_d) H^{(N)} \ket{0}
   \end{eqnarray}

Here, $x_j$ is the $j$-th component of the input vector ${\bf x}$, $\overline{S}_3$ is the diagonal matrix $diag (-(N-1)/2, \cdots,(N-1)/2)$, $\ket{0} = (1,0,0, \cdots, 0)^\dag$ and $H^{(N)}$ is the generalized Hadamard gate which can be determined from following set of equations:

\begin{equation}
    \label{genhadamard}
    H^{(N)} \ket{j} = \frac{1}{\sqrt{N}} \sum_{k=0}^{N-1} e^{i \frac{2 \pi j k}{N}} \ket{k} ; \ \ j=0,1, \cdots, N-1
\end{equation}

 For $N=2$, $H^{(N)}$ reduces to the standard Hadamard gate. Variables $\{w_i; i=1,2, \cdots, d\}$ are of special interest. These are free parameters that can be tuned, in the course of training. Clearly this is a parameter dependent state preparation procedure. This is not a hindrance however, as will be explained in Sec.~\ref{disc}.

\subsection{Parameterized unitary operations}
\label{puo}

Several combinations of fundamental gates have been proposed so far to carry out  the unitary operation in conventional multi-qubit based quantum classifiers. \cite{havliv19,Farhi18,Schuld18} Nevertheless there can be an infinitely large number of such combinations and there exists no unique way to determine which one has to be chosen. Certainly, the most general unitary operation, in this case, is not known.

Our model, on the other hand, bypasses this ambiguity. Input vectors ${\bf x}$ are encoded in states of a single qu$N$it. The most general unitary operation that can be applied to such a state is known. It is an element of the $SU(N)$ group. It admits a generalised Euler angle parameterisation as:

\begin{equation}
   \label{group}
    SU(N)(\{\alpha_j\}) = \Big( \prod_{2 \leq k \leq N} A(k) \Big) SU(N-1) e^{i \lambda_{(N^2 -1)} \alpha_{(N^2 -1)}}
\end{equation}

with $ A(k) = e^{i \lambda_3 \alpha_{(2k-3)}} e^{i \lambda_{((k-1)^2+1)} \alpha_{(2(k-1))}}$ and $SU(1) = 1$. The set $\{ \alpha_j; j = 1, 2, \cdots, N^2-1)\}$ are the learnable parameters while $\{ \lambda_j\}$ constitute the Lie algebra of $SU(N)$ (for further details see \cite{Tilma02}).

For a simple demonstration, consider the case where $N=2$. The unitary operation is therefore an element of the $SU(2)$ group and admits an expansion $e^{i \lambda_3 \alpha_{1}}  e^{i \lambda_{2} \alpha_{2}} e^{i \lambda_{3} \alpha_{3}} $. $\lambda_1,\lambda_2$ and $\lambda_3$ are the Pauli spin matrices, which are generators of rotation and hence constitutes the Lie Algebra of $SU(2)$. Thus the operation can be thought of as a series of consecutive rotations: rotation by angle $\alpha_{3}$ about z axis, followed by rotation by angle $\alpha_{2}$ about y axis and then again rotation by angle $\alpha_{1}$  about z axis.

\subsection{Measurement and decision functions}
\label{measurement}
The final stage of the model is where the classifier predicts the label for an input vector. In our case this decision making is determined by a projective measurement on the transformed state $ \ket{\tilde{\psi}(\bf x)} = SU(N)  \ket{\psi(\bf x)}$. We typically measure the $S_3$ operator, though any other non-degenerate operator would do the job equally well; $S_3 = Diag(1,2, \cdots, N)$. 

In order to predict the label associated with ${\bf x}$ we have to first construct the complete set of probabilities $\{ p_a \}$, corresponding to all outcomes of $S_3$, for the state $\ket{\tilde{\psi}(\bf x)}$.  $p_a = Tr (\pi_a \ket{\tilde{\psi}(\bf x)} \bra{\tilde{\psi}(\bf x)})$ is the probability of the outcome $S_3 = a$; $S_3 = \sum_{a=0}^N a \pi_a$. We assert the rule:

\begin{equation}
    p_b = max\{ p_a \} \rightarrow f^*({\bf x}) = l_b
\end{equation}

,i.e., the predicted label will be $l_b$ iff $S_3 = b$ is the most likely outcome of the measurement. The predicted label is not necessarily the same as actual label. Hence to minimize mismatches we need to train the circuit. We say that a sample ${\bf x}$ is correctly classified iff $f^*({\bf x}) = f({\bf x})$, as explained in Sec. ~\ref{meth}. .

The forward part of the computation constitutes of three stages, as discussed in Sec.~\ref{sp}, \ref{puo} and \ref{measurement}. In a classical FCNN, the input vectors are multiplied, in the forward computing  part, with weight matrices. This step is popularly known as Vector Matrix Multiplication (VMM) operation. Each element of the weight matrix is an adjustable parameter in the network and can be updated during training  \cite{LeCun}. In contrast, the trainable parameters in our circuit are introduced through a series of unitary operations. In essence this is also a VMM operation.

\subsection{Training}
\label{training}

Consider a training set $\mathcal{T} \in \mathcal{S}$. We assume that there are $m_k$ input vectors in $\mathcal{T}$ with the label $l_k$. We start with the encoding ${\bf x} \rightarrow \ket{\psi(\bf x)}$ (see Sec.~\ref{sp}). The training procedure that we employ is what we call - ``single shot training". The idea is to use all input states (samples) that belong to the same class, simultaneously, to train the circuit.   This gives a speedup to the training procedure, as opposed to the more conventional mean where one trains using a single state / input sample at a time. To distinguish between states  belonging to different classes, we employ a new notation.  We shall denote the states that are associated with label $l_k$ as $\{\ket{\psi(\bf x)}^k_i ; i = 1, \cdots, m_k\}$. $m_k$ is the total number of such states.   We are interested in the ensemble $\{(\ket{\psi(\bf x)}^k_i, 1/m_k)\}$. It constitutes of all states belonging to the same class associated with label $l_k$, each occurring with equal probability $1/m_k$ in the ensemble. In quantum mechanics such a quantity is represented by the density matrix:

\begin{equation}
    \rho_k = \frac{1}{m_k} \sum_{i=1}^{m_k} \ket{\psi(\bf x)}^k_i \bra{\psi(\bf x)}^k_i
\end{equation}

$\rho_k$ is then allowed to pass through the quantum circuit, where it first gets transformed as $\tilde{\rho}_k = SU(N) \rho_k SU(N)^\dag$, as described in Sec. ~\ref{puo}. Then  measurement is carried out on the transformed state, as described in Sec. ~\ref{measurement}.

For a state with label $l_k$ to be correctly classified, probability $p_k$ needs to be higher than probabilities $\{p_{k'}; k' = 1, 2,  \cdots, N,  k' \neq k\} $. A stronger, or more ideal assertion  requires $p_k$ to be 1. Hence, we consider the error function for the $k$-th class to be:  $E_k = 1 - p_k$. We repeat the exercise for all classes to finally arrive at the total error:

\begin{equation}
    {\bf E} = \frac{1}{M} \sum_{k=1}^N m_k E_k
\end{equation}

where $M = \sum_{k=1}^N m_k$ is the total number of samples in the training set $\mathcal{T}$. ${\bf E}$ is simply the weighted sum of the errors for each class. Evaluation of ${\bf E}$ completes one iteration or epoch.

The objective of the training process is to minimize the error function ${\bf E}$, which can  be carried out on a classical computer since ${\bf E}$ is a classical function.  After each epoch, the training parameters $\{ w_i; i=1, 2, \cdots, d\}$ and $\{\alpha_i; i=1, 2, \cdots, (N^2-1)\}$ are updated to minimize ${\bf E}$ using the classical gradient descent technique \cite{LeCun}. This is the feedback part of the computation. After that, the forward part of the computation is carried out once again. This process is repeated till the value of ${\bf E}$ converges to a minimum. At that stage the values of the free parameters are frozen and training is complete.

It is to be noted that in classical machine learning, every sample in the training set is trained separately in each epoch and then again the same training is repeated for all samples with updated parameters in the next epoch \cite{LeCun}. On the other hand, all samples belonging to the same class are trained at once here. Once each class is trained, one epoch is completed.

\begin{figure}
  \centering
  \subfloat[]{\includegraphics[width=0.5\textwidth]{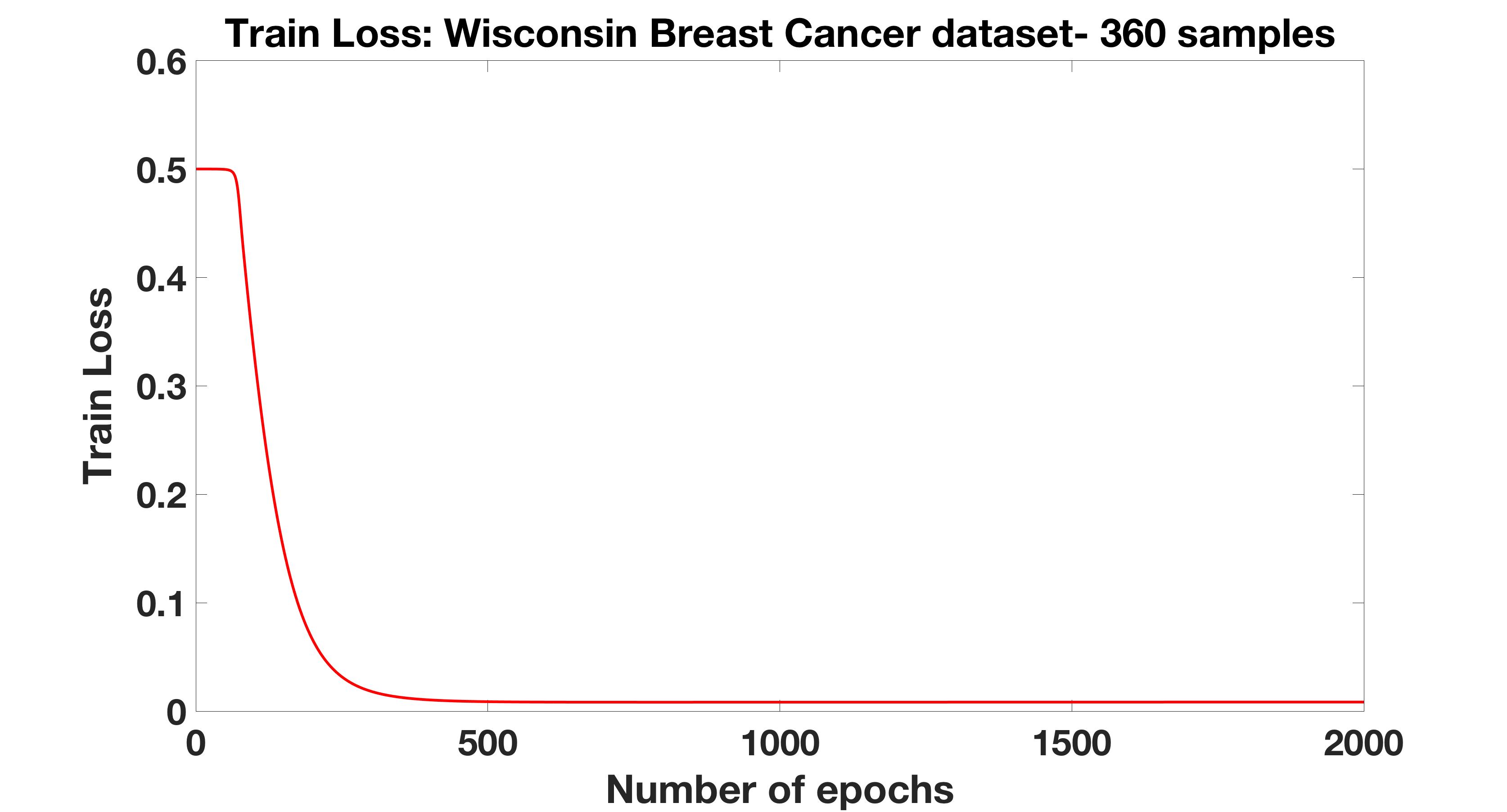}\label{trWBCE}}
  \hfill
  \subfloat[]{\includegraphics[width=0.5\textwidth]{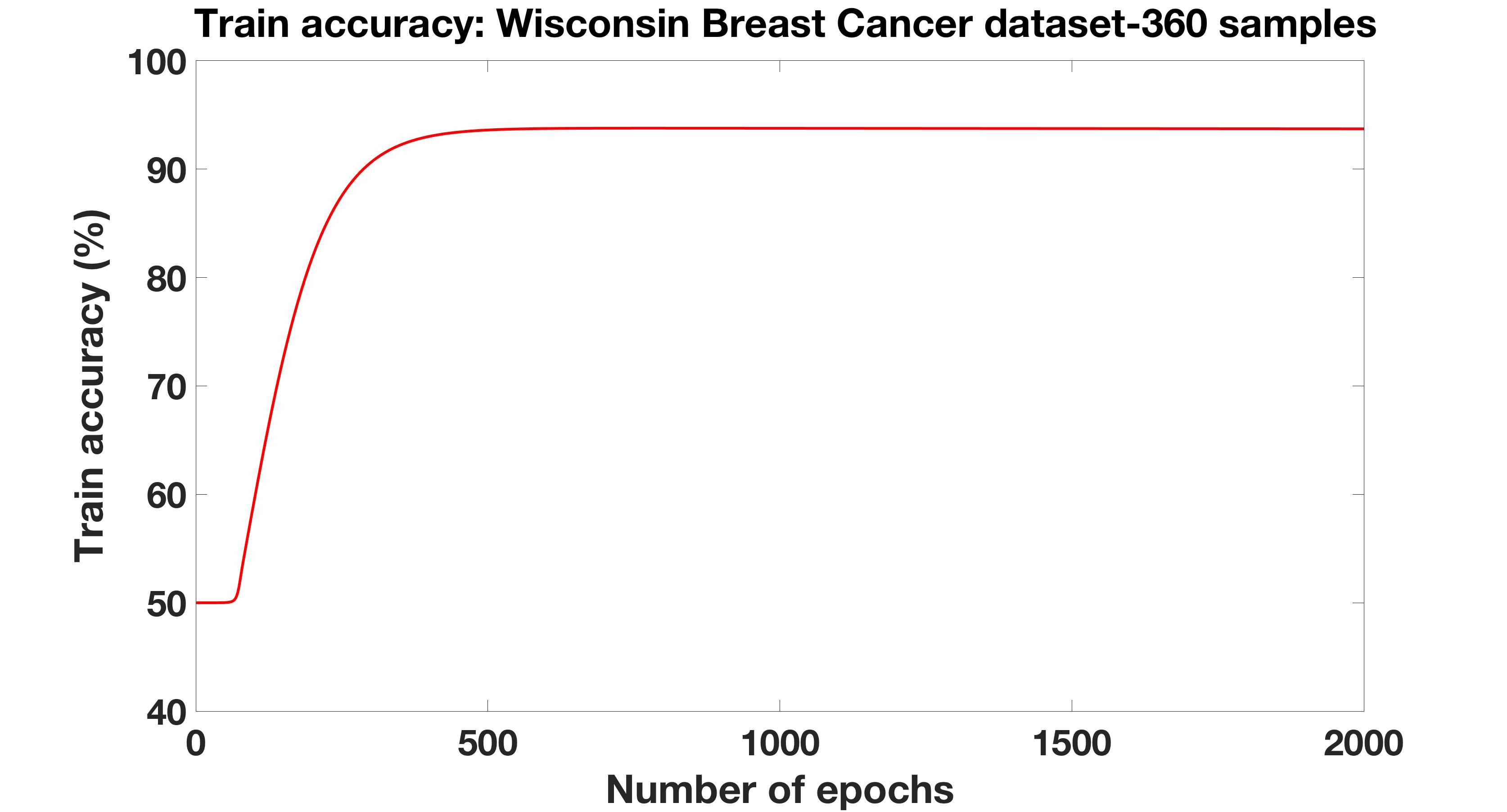}\label{trWBCA}}
  \caption{\label{traintesterr1} Training results for CANCER dataset. (a) Variation of error with number of epochs. (b) Variation of train accuracy with number of epochs.}
\end{figure}

\section{Results}
\label{results}
To demonstrate how effective our quantum classifier is, we present training results for four benchmark datasets. We shall also compare the performance of our quantum classifier with that of classical FCNN on the same datasets. The datasets that we have chosen are CANCER (WBC- \cite{WBC}), SONAR \cite{Sonar}, IRIS (Fisher's Iris- \cite{FisherIris}) and IRIS(2). IRIS(2) is a modified IRIS dataset consisting of only two linearly separable classes (``setosa" and ``virginica").  The essential features of the datasets are shown in Table~\ref{dandN}.

\begin{table}[ht!]
\begin{center}
\begin{tabular}{| l | l | l |}  
\hline
{\bf Dataset} & $d$ & $N$ \\ \hline
CANCER & 27 & 2 \\ \hline
SONAR & 60 & 2 \\ \hline
IRIS & 4 & 3 \\ \hline
IRIS(2) & 4 & 2 \\ \hline
\end{tabular}
\end{center}
\caption{\label{dandN} The dimension of the input vectors ($d$) and number of classes in different datasets ($N$)}
\end{table}

None of these datasets are pre-processed. More information on the datasets can be found in \cite{FisherIris,WBC,Sonar}. 

\begin{figure}
  \centering
  \subfloat[]{\includegraphics[width=0.5\textwidth]{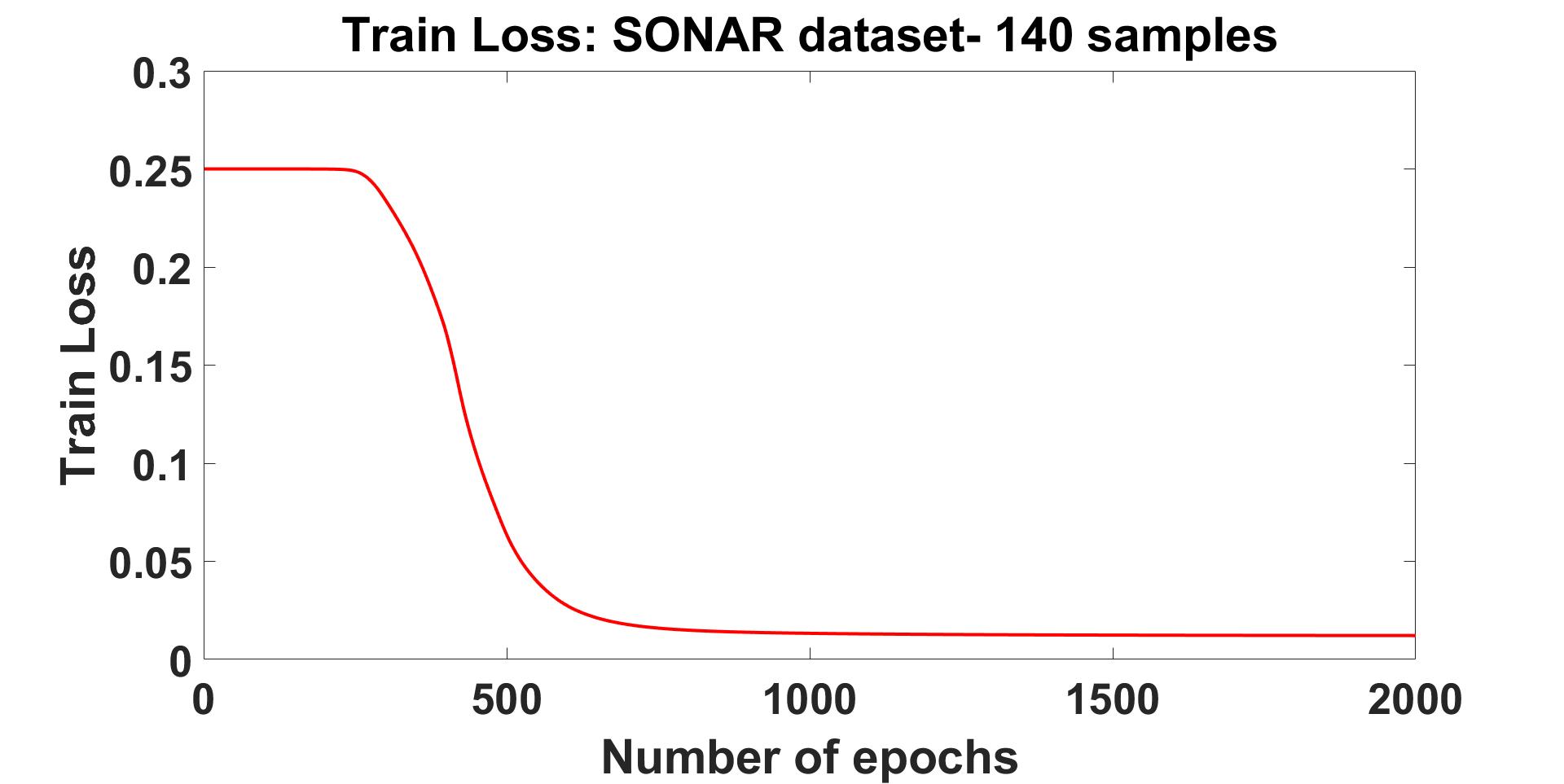}\label{trSONE}}
  \hfill
  \subfloat[]{\includegraphics[width=0.5\textwidth]{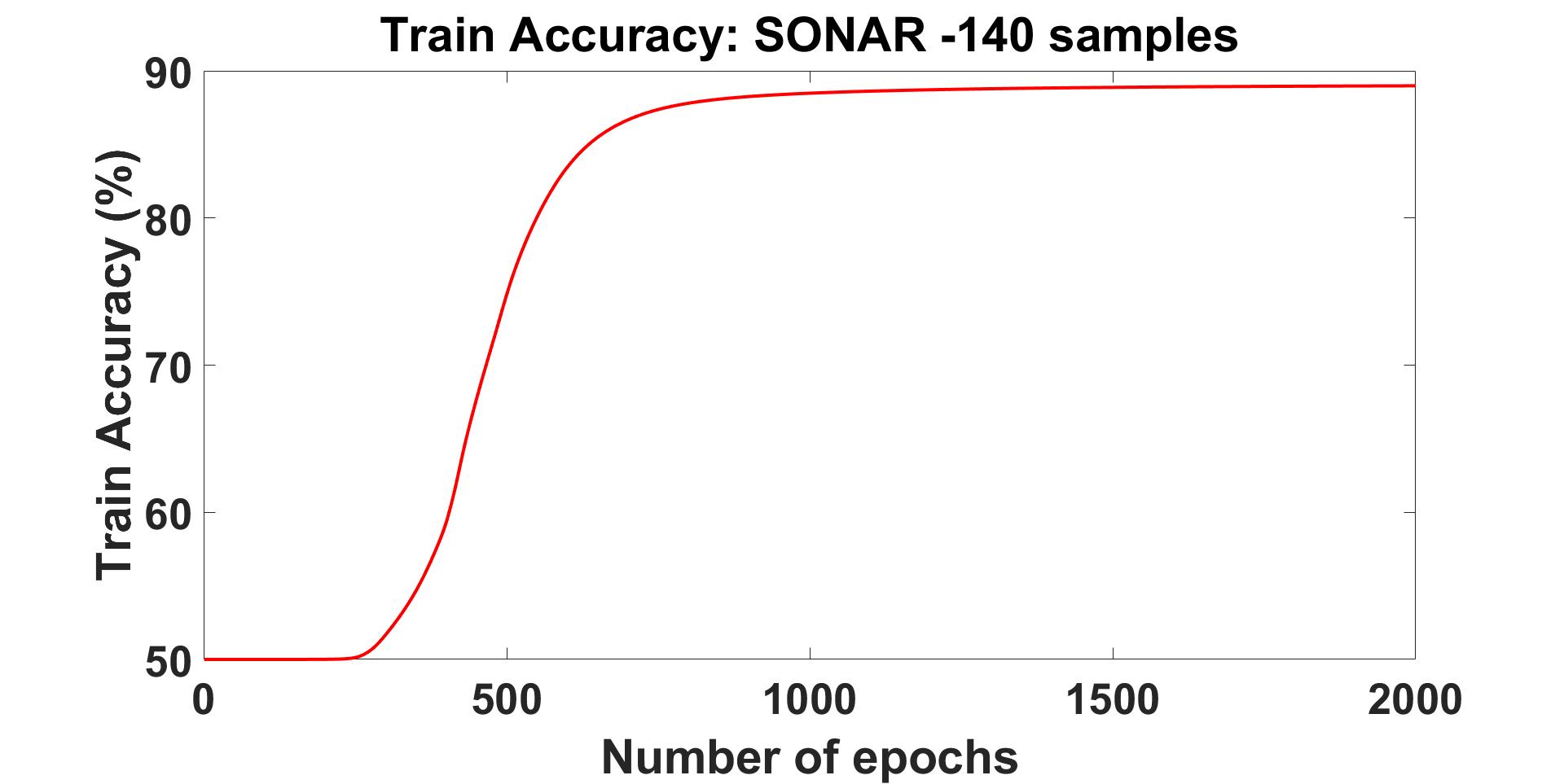}\label{trSONA}}
  \caption{\label{traintesterr2} Training results for SONAR dataset. (a) Variation of error with number of epochs. (b) Variation of train accuracy with number of epochs.}
\end{figure}

\begin{figure}
  \centering
  \subfloat[]{\includegraphics[width=0.5\textwidth]{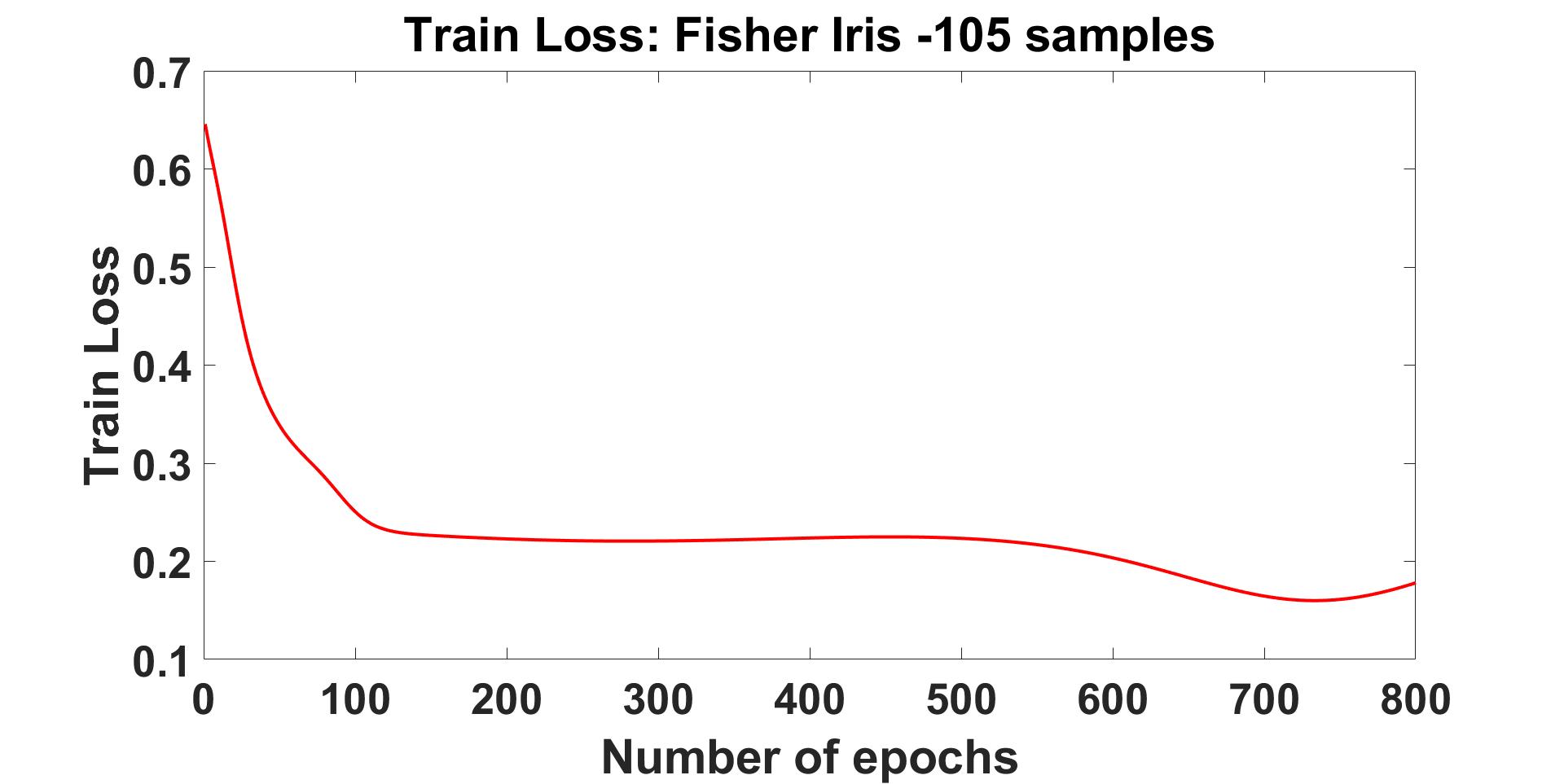}\label{trIRISE}}
  \hfill
  \subfloat[]{\includegraphics[width=0.5\textwidth]{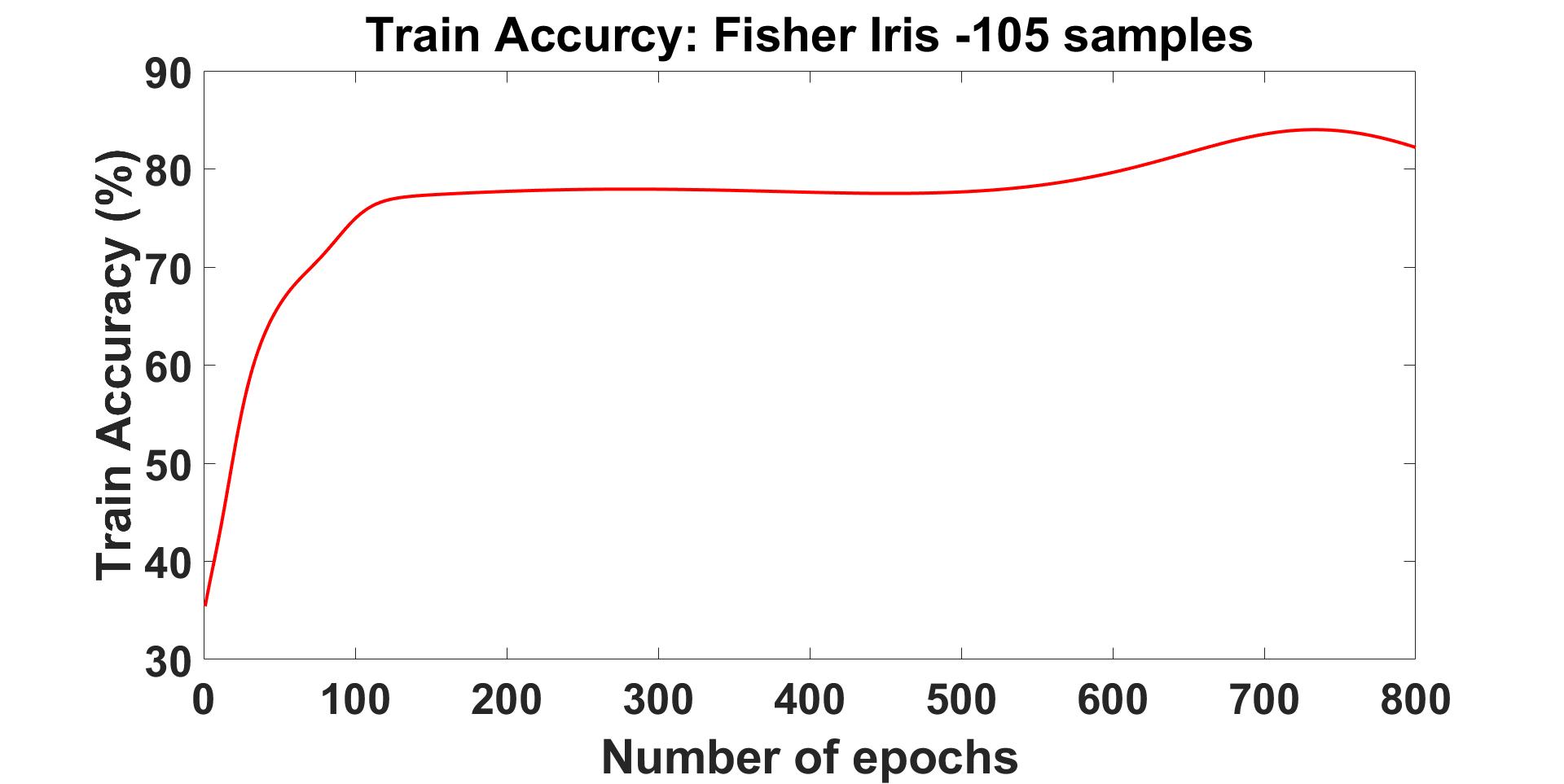}\label{trIRISA}}
   \caption{\label{traintesterr3} Training results for IRIS dataset. (a) Variation of error with number of epochs. (b) Variation of train accuracy with number of epochs. }
\end{figure}

\begin{figure}
  \centering
  \subfloat[]{\includegraphics[width=0.5\textwidth]{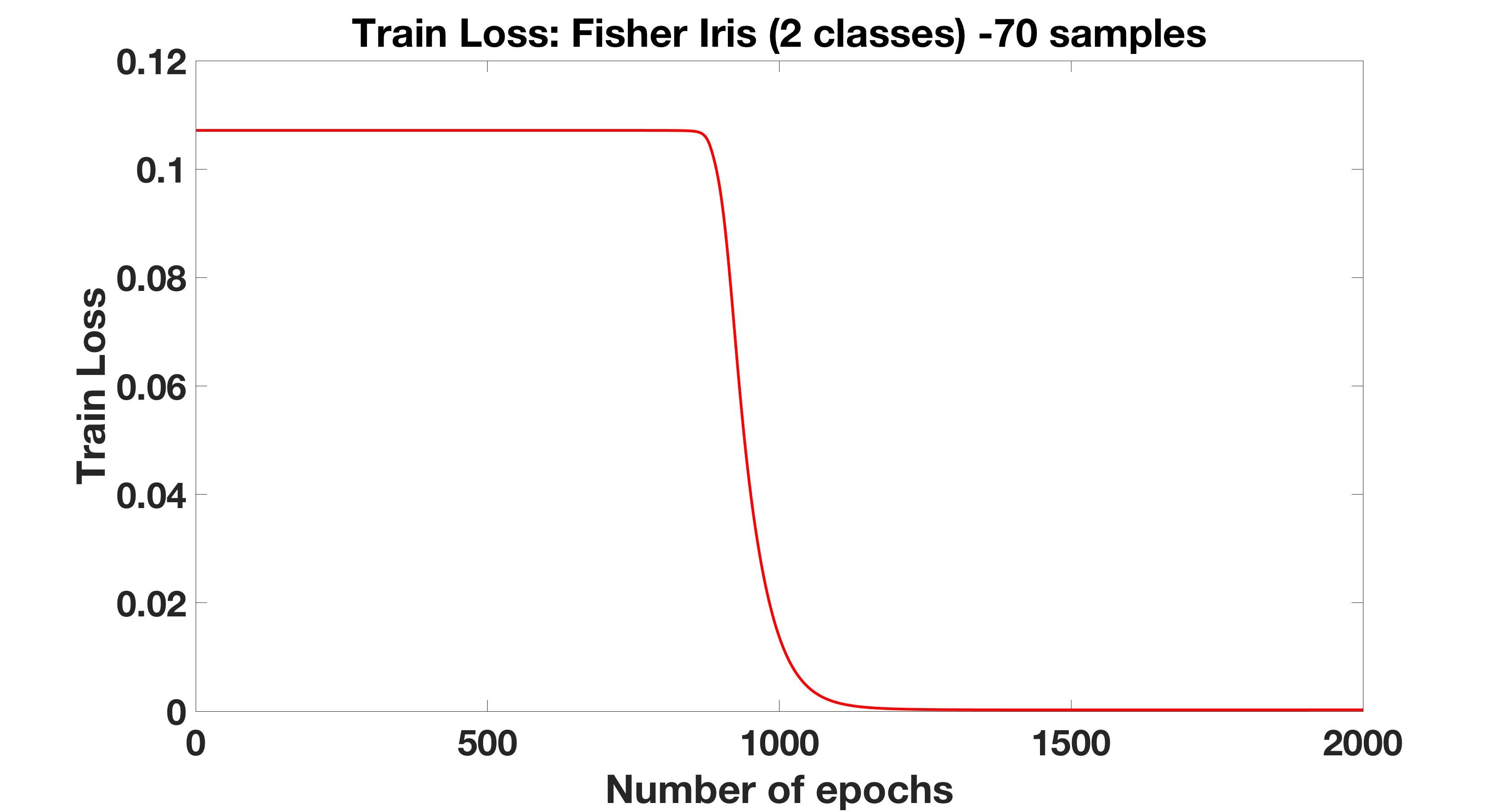}\label{trIRISE2}}
  \hfill
  \subfloat[]{\includegraphics[width=0.5\textwidth]{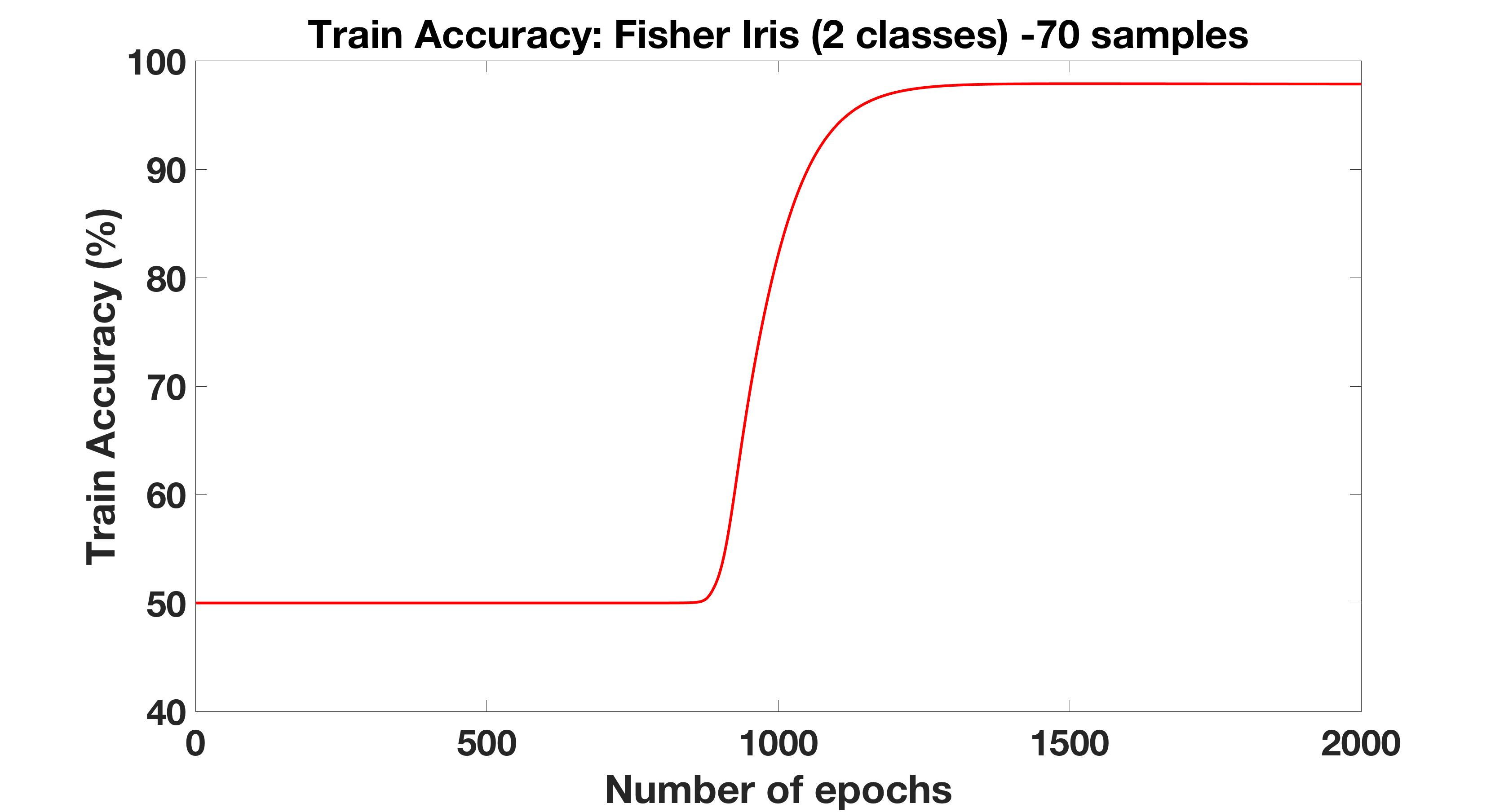}\label{trIRISA2}}
  \caption{\label{traintesterr4} Training results for IRIS(2) dataset. (a) Variation of error with number of epochs. (b) Variation of train accuracy with number of epochs.}
\end{figure}

To train the circuit we choose  a subset $\mathcal{T}$ from each dataset with the cardinality $n( \mathcal{T} ) = 360, 140$, $105$ and $70$ for  CANCER, SONAR, IRIS and IRIS(2) respectively. We ensure that for a given $\mathcal{T}$, all classes are represented equally, i.e, $m_i = m_j  \forall  (i,j)$. The learning rate is kept fixed at 0.002 at all time. Fig.~\ref{traintesterr1}, \ref{traintesterr2}, \ref{traintesterr3}, \ref{traintesterr4} shows how the error converges with the number of epochs. The train accuracy can be seen to behave complementary to the error, as is expected. Train accuracy is calculated as the percentage of train samples that are correctly classified. As stated in Sec.~\ref{training}, the numerical values of the training parameters are frozen, once the error reaches its minimum. The final train and test accuracy are calculated at this stage. The results are tabulated in Table~\ref{qnn}.

\begin{table}[ht!]
\begin{center}
\begin{tabular}{| l | l | l |}  
\hline
{\bf Dataset} & {\bf Train accuracy} & {\bf Test accuracy} \\ \hline
CANCER & 93.71 $\pm$ 0.316$\%$ & 90.19 $\pm$ 0.3119$\%$ \\ \hline
SONAR & 85.34 $\pm$ 3.863$\%$ & 41.25 $\pm$ 4.863$\%$ \\ \hline
IRIS & 81.99 $\pm$ 1.459$\%$ & 83.26 $\pm$ 1.571$\%$ \\ \hline
IRIS(2) & 97.89 $\pm$ 0.04$\%$ & 96.96 $\pm$ 0.21$\%$ \\ \hline
\end{tabular}
\end{center}
\caption{\label{qnn} Mean Train and test accuracy for our chosen datasets as obtained from our quantum classifier. For each dataset, the classifier was run for 100 times.}
\end{table}

The corresponding results for training using a classical FCNN is shown in table~\ref{cnn}. The network has just two layers - input and output and uses $tanh$ activation function. No hidden layers were used. We maintained the learning rate at 0.1 during the entire process.

\begin{table}[ht!]
\begin{center}
\begin{tabular}{| l | l | l |}  
\hline
{\bf Dataset} & {\bf Train accuracy}  & {\bf Test accuracy}  \\ \hline
CANCER & $96.38 \pm 0\%$ & $92.5 \pm 0\%$ \\ \hline
SONAR & $95 \pm 1.74\%$ & $44 \pm 2.54\%$ \\ \hline
IRIS & $66.66 \pm 0\%$ & $65 \pm 0\%$ \\ \hline
IRIS(2) & $100 \pm 0.0\%$ & $100 \pm 0.0\%$ \\ \hline
\end{tabular}
\end{center}
\caption{\label{cnn} Mean train and test accuracy as obtained from a classical neural network with zero hidden layer. Considering the fact that classical neural networks converge efficiently, the classifier was run for just 5 times per dataset.}
\end{table}

Based on the results in tables~\ref{qnn} and \ref{cnn}, we find that the accuracy numbers from our quantum classifier and classical FCNN with no hidden layer are comparable for CANCER, SONAR and IRIS(2) datasets. For the IRIS dataset our classifier outperforms its classical counterpart.

In the IRIS dataset, which is the original 3 output class dataset provided in \cite{FisherIris}, samples of the ``setosa" class are linearly separable from that of the ``virginica"  and ``versicolor"  classes. Hence classical FCNN with no hidden layer can distinguish between ``setosa"  and ``versicolor" or between  ``setosa" and ``virginica"  , but cannot distinguish between ``virginica" and ``versicolor" \cite{FisherIris}. As a result, maximum accuracy on the IRIS dataset with classical FCNN is 66 $\%$ (Table ~\ref{cnn}). However, our quantum classifier classifies IRIS with accuracy higher than 80 $\%$ (Table ~\ref{qnn}). The IRIS(2) dataset is prepared by only taking 2 classes, which are linearly separable, from the total 3 in original Fisher's Iris dataset. Hence, classical FCNN without a hidden layer has 100 $\%$ classification accuracy on IRIS(2) dataset (Table ~\ref{cnn}). Our quantum classifier also shows comparable performance (Table ~\ref{qnn}).

\begin{table}[ht!]
\begin{center}
\begin{tabular}{| l | l | l |}  
\hline
{\bf Dataset} & {\bf Train accuracy}  & {\bf Test accuracy} \\ \hline
CANCER & $98.71 \pm 0.228\%$ & $94.82 \pm 0.82\%$ \\ \hline
SONAR &  $97 \pm 0.7\%$ &  $59.5 \pm 3\%$ \\ \hline
IRIS & $96.88 \pm 2.58\%$ & $99.5 \pm 1.11\%$ \\ \hline
IRIS(2) & $100 \pm 0.0\%$ & $100 \pm 0.0\%$ \\ \hline
\end{tabular}
\end{center}
\caption{\label{cnn1} Mean train and test accuracy as obtained from a classical neural network with one hidden layer. Considering the fact that classical neural networks converge efficiently, the classifier was run for just 5 times per dataset.}
\end{table}

The situation with classical FCNN however changes once hidden layers are introduced. Table~\ref{cnn1} show results for a FCNN having a single hidden layer. The dimension of the hidden layer has been adjusted for highest accuracy given each dataset. We have chosen them to be 10, 30, 4 and 4 for CANCER, SONAR, IRIS and IRIS(2) respectively. The activation function and the learning rate are the same as it was earlier. After having introduced a hidden layer (of suitable dimension), a neural network can perfectly classify even  linearly inseparable datasets as evident from the very high classification accuracy values in Table ~\ref{cnn1}. This result follows from the universal approximation theorem \cite{Hornik89}. The quantum classifier proposed here, however can be seen, not to have achieved this standard as of now.

\section{Discussion and conclusion}
\label{disc}

In conclusion, we have proposed a quantum classifier using a variational quantum circuit and used it to classify four benchmark datasets {\it viz.} CANCER, SONAR, IRIS and IRIS(2). Among these datasets, we observe that it can classify the linearly separable ones completely.For the linearly inseparable datasets, it performs better than a classical FCNN with no hidden layer (Table-\ref{qnn},\ref{cnn}). 

The simplest and also the most apparent feature of our classifier is the use of a single qu$N$it at the input layer. The dimension of the corresponding Hilbert space is also not too high as it is determined by the number of classes ($N$) as opposed to the dimension of the input vectors ($d$). It is well known that $d >> N$ for most cases. This is in stark contrast to both classical NN and most of the other quantum classifiers. The input layer of the former typically require $d$ nodes while the input layer of the latter is made up of multiple qubits, whose number is also determined by the dimension of the input vectors \cite{havliv19,Schuld19,Farhi18}.

Also, for the datasets we have used, there is a significant reduction in the number of  tunable training parameters compared to classical NN. A simple calculation shows that we need just 
$(d+N^2-1)$ parameters in our classifier, the parameters being: $\{ w_i; i = 1, 2, \cdots, d\}$ and  $\{ \alpha_i; i = 1, 2, \cdots, (N^2-1)\}$ . However for classical FCNN, without a hidden layer, the number of weight parameters sums up to $(N+1)d$, which is already
much larger than that for our quantum classifier.

The subsequent addition of hidden layers to the classical FCNN increases the number of training parameters in it even further. Nevertheless, introduction of a hidden layer has been shown to outperform our classifier (Table-\ref{qnn},\ref{cnn1}) . This reflects that our classifier can not resolve linearly inseparable classes while the latter can. Though a few steps has been taken in this direction \cite{havliv19,Schuld19}, to the best of our knowledge the most general quantum algorithm that can classify all linearly inseparable datasets with $100\%$ accuracy is yet to be known. The larger problem, is to identify the set of encodings ${\bf x} \rightarrow \ket{\psi(\bf x)}$ such that a linearly inseparable set $\mathcal{S}$ gets mapped onto a linearly separable one in the quantum feature space. In principle, this is possible, as the encoding ${\bf x} \rightarrow \ket{\psi(\bf x)}$ is itself a nonlinear operation. 


From a more physics point of view this works brings to the fore that quantum correlations are not imperative, as resources, to quantum machine learning algorithms. Our classifier uses the simple property of superposition and can be seen to classify labelled data with good efficiency.

Finally, before concluding, we will like to comment on our training procedure and its associated subtleties. Since our algorithm employs projective measurements to generate the decision functions, it necessitates the states to be prepared freshly after each epoch, with updated gate parameters. This is true for all quantum classifiers that involve projective measurements and therefore  is not a shortcoming of our classifier only.

\section*{Acknowledgement}
SA thanks CSIR for funding his research (Grant no. - 09/086(1203)/2014-EMR-I)

\section*{References}

\bibliographystyle{iopart-num}  
\bibliography{bibliography}

\end{document}